\documentclass[conference]{IEEEtran}
\IEEEoverridecommandlockouts
\usepackage{cite}
\usepackage{amsmath,amssymb,amsfonts}
\usepackage{algorithmic}
\usepackage{graphicx}
\usepackage{textcomp}
\usepackage{xcolor}
\usepackage{url}
\usepackage{subfigure}
\usepackage{float}
\usepackage{bm}
\usepackage{cases} 
\usepackage[top=0.55in, bottom=0.55in, left=0.635in, right=0.635in]{geometry}
\def\BibTeX{{\rm B\kern-.05em{\sc i\kern-.025em b}\kern-.08em
    T\kern-.1667em\lower.7ex\hbox{E}\kern-.125emX}}
\begin{document}

\title{Optimizing Information Freshness in Two-Way Relay Networks}
	
%	{\thanks{This work is done when the first author Bohai Li is a visiting student at the Chinese University of Hong Kong.}
%	}
%	{\thanks{This work of Zheng Dong was supported by the Fundamental Research Funds of Shandong University (61170079614095).}
%	}

\author{\IEEEauthorblockN{Bohai Li\textsuperscript{1}, He Chen\textsuperscript{2}, Nikolaos Pappas\textsuperscript{3} and Yonghui Li\textsuperscript{1}}%
	\IEEEauthorblockA{\textsuperscript{1}School of Electrical and Information Engineering, The University of Sydney, Sydney, Australia}
	\IEEEauthorblockA{\textsuperscript{2}Department of Information Engineering, The Chinese University of Hong Kong, Hong Kong SAR, China}
	\IEEEauthorblockA{\textsuperscript{3}Department of Science and Technology, Link\" oping University, Campus Norrk\" oping, Sweden\\	\textsuperscript{1}\{bohai.li, yonghui.li\}@sydney.edu.au,
		\textsuperscript{2}he.chen@ie.cuhk.edu.hk, \textsuperscript{3}nikolaos.pappas@liu.se}
	
}

\maketitle

\begin{abstract} 
	
In this paper, we investigate an amplify-and-forward (AF) based two-way cooperative status update system, where two sources aim to exchange status updates with each other as timely as possible with the help of a relay. Specifically, the relay receives the sum signal from the two sources in one time slot, and then amplifies and forwards the received signal to both the sources in the next time slot. We adopt a recently proposed concept, the age of information (AoI), to characterize the timeliness of the status updates. Assuming that the two sources are able to generate status updates at the beginning of each time slot (i.e., \textit{generate-at-will} model), we derive a closed-form expression of the expected weighted sum AoI of the considered system. We further minimize the expected weighted sum AoI by optimizing the transmission power at each node under the peak power constraints. Simulation results corroborate the correctness of our theoretical analysis.

\end{abstract}

\section{Introduction}
With the abundance of inexpensive Internet of Things (IoT) devices, there has been a rapid development in real-time monitoring applications, such as autonomous vehicles, wireless industrial automation, and health care monitoring \cite{b1}, \cite{b2}. In these applications, timely delivery of status updates is crucial. To quantify the timeliness of status updates, a new performance metric, termed the age of information (AoI), was recently introduced in \cite{b3}. The AoI is defined as the time elapsed since the generation of the latest received status update at the destination. According to the definition, unlike any conventional metrics, the AoI is jointly determined by the transmission interval and the transmission delay \cite{b4}. As a result, the AoI has attracted increasing attention as a more comprehensive evaluation criterion for information freshness \cite{b1, b2, b3, b4, b5, b6, b7, b8, b9, b10, b11, b12, b13, b14, b15, b16, b17}.

Since the AoI concept was first proposed to characterize the information freshness in a vehicular status update system \cite{b5}, AoI-oriented problems have been well studied in single-hop networks \cite{b6, b7, b8, b9, b10, b11, b12, b13}. However, only a handful of work considered multi-hop networks \cite{b14, b15, b16, b17}. In \cite{b14}, both the offline and online policies were proposed to minimize the AoI of a two-hop energy harvesting network. The authors in \cite{b15} proved that the preemptive Last Generated First Served (PLGFS) queuing policy is age-optimal in a multi-hop network, where a single source disseminates status updates through a gateway to the network. Both \cite{b16} and \cite{b17} considered the AoI in multi-source, multi-monitor, multi-hop networks, while \cite{b17} was from a \textit{global} perspective in the sense that every node in the network is both a source and a monitor. All the above work assumed that only one node can transmit in each time slot such that the transmission is always one-way. However, in the applications where two nodes sometimes need to exchange status updates with each other through a single relay (e.g., autonomous vehicles), one-way transmission means that one node must wait while the other node is transmitting. In such multi-hop networks, the two-way relaying scheme, in which the nodes simultaneously transmits status updates to each other, can significantly decrease the AoI in the networks. To the best of the authors' knowledge, although the two-way relaying scheme can potentially enhance the information freshness in the multi-hop network, there is no existing work that analyzes and optimizes the AoI of a two-way cooperative status update system. Such an analysis and optimization is non-trivial because the AoI of the two source nodes are tangled together. Specifically, transmitting with higher power at one source node reduces the outage probability and the AoI at the corresponding destination. However, higher transmission power at this source node also means that the interference in its received signal will be stronger, which results in a larger local AoI.

%However, {\color{red}this is not clear}at this source node, the interference in the received signal will be stronger, which results in a larger AoI.

%the two-way relaying, in which the nodes can, the one-way transmission 
%
%This scheme obviously increases the total age in the multi-hop network compared to the two-way relaying, in which the two nodes simultaneously transmit status update to each other.
%
%Comparing with the two-way relaying, in which the two nodes simultaneously transmits status updates to each other, one-way scheme obviously increase the total AoI of the nodes in the multi-hop network.

Motivated by this gap, in this paper we investigate an amplify-and-forward (AF) based two-way cooperative status update system, in which two source nodes timely exchange status updates with each other with the help of a relay. In order to keep the information at each destination as fresh as possible, we adopt the two time slot physical-layer network coding (PNC) scheme and the \textit{generate-at-will} model in the system. We concentrate on analyzing and optimizing the AoI of the considered system. We first define some necessary time intervals to mathematically express the average AoI at each destination node. By representing these time-interval definitions in terms of the transmission success probabilities, we then attain a closed-form expression of the expected weighted sum AoI for the considered system. As the transmission success probabilities are functions of the transmission power at each node, we further minimize the expected weighted sum AoI by optimizing the transmission power at each node under the peak power constraints. Simulation results are then provided to validate the theoretical analysis.

\section{System Model and Scheme Description}
\subsection{System Model}
We consider a two-way cooperative status update system where two source nodes, source $A$ ($S_A$) and source $B$ ($S_B$), want to exchange status updates with each other as timely as possible with the help of a single relay ($R$). $R$ adopts the amplify-and-forward (AF) relaying protocol. Specifically, $S_A$ and $S_B$ transmits status updates $x_A$ and $x_B$, which can be generated at the beginning of any time slot (i.e. \textit{generate-at-will} model\cite{b18}), to $R$ with power $P_A$ and $P_B$, respectively. Since the AF protocol is adopted, the received signals at $R$ from $S_A$ and $S_B$ are multiplied by a gain $G$. The amplified signal is then 
forwarded to the source nodes with the power $P_r$. To quantify the timeliness of the status updates, we adopt a recently proposed AoI metric, which is defined as the time elapsed since the generation of the last successfully received status update. The considered system is time slotted, and the transmission of each status update by any node takes exactly one time slot. The length of each time slot is normalized to one without loss of generality.  
%$S_A$/$S_B$/$R$

We denote by $h_A$ and $h_B$ the channel coefficients between $S_A$ and $R$, and $S_B$ and $R$, respectively. All channels in the system are considered to be subject to Rayleigh fading. We assume that the mean and the variance associated with the Rayleigh distribution are 0 and 1, respectively, such that $h_A, h_B \sim \mathcal{CN}(0,1)$. In addition, we denote $n_A \sim \mathcal{CN}(0,\sigma_A^2)$, $n_B \sim \mathcal{CN}(0,\sigma_B^2)$ and $n_r \sim \mathcal{CN}(0,\sigma_r^2)$ as the additive white Gaussian noise (AWGN) at $S_A$, $S_B$ and $R$, respectively. All the channels are assumed to be constant during one round of status update exchange between $S_A$ and $S_B$\cite{b19}. With the aid of the instantaneous channel state information (CSI) and the noise statistics, the gain $G$ can be properly chosen such that the instantaneous transmission power at $R$ is constrained\cite{b20}. Furthermore, $G \sqrt{P_r}$ is assumed to be known at both source nodes. For notation simplification, we define $\gamma_{A}=\frac{P_{A}}{\sigma_r^2}$, $\gamma_{B}=\frac{P_{B}}{\sigma_r^2}$, $\gamma_{r,A}=\frac{P_{r}}{\sigma_A^2}$ and $\gamma_{r,B}=\frac{P_{r}}{\sigma_B^2}$.

%We consider that all channels in the system are subject to Rayleigh fading. The channel coefficients between $S_A$ and $R$, and $S_B$ and $R$ are denoted by {\color{red}$h_A \sim \mathcal{CN}(0,1)$ and $h_B \sim \mathcal{CN}(0,1)$, respectively.}

%three transmission schemes, which differ in the number of time slots used for $S_A$ and $S_B$ to communicate with each other. Specifically, in these three transmission schemes, $S_A$ and $S_B$ completes one round of information exchange over either two, three or four time slots, as shown in Fig. 1. We now start by describing the transmission schemes.

%one round of information exchange between $S_A$ and $S_B$. Specifically, $S_A$ and $S_B$ communicates with each other over either two, three or four time slots, as shown in Fig. 1. We now start by de

%We also assume that the channels associated with each node are known at that node, i.e., $h$ is known at $S_A$, $g$ is known at $S_B$, and both $h$ and $g$ are known at $R$.

%{\color{red}We assume that the channels and noise are constant during these time slots}, and that each source knows the instantaneous CSI of its associated channel, i.e., $h$ is known at $S_A$ and $g$ is known at $S_B$. 

\subsection{Two Time Slot PNC Scheme}

In this paper, we consider the two time slot PNC scheme, in which $S_A$ and $S_B$ completes one round of status update exchange over two time slots, as shown in Fig. 1. In the first time slot, the two source nodes simultaneously transmit to $R$. In the second time slot, $R$ amplifies and forwards the received signals to the two sources nodes to complete the status update exchange.

Based on \cite[Eqs. (8) and (9)]{b19}, we can express the received signal-to-noise ratio (SNR) at $S_{\zeta}$, given by
\begin{equation}
\Gamma_{{\zeta},2TS}=\frac{\gamma_{r,{\zeta}}\gamma_{\varrho}|h_{\zeta}|^2|h_{\varrho}|^2}{(\gamma_{r,{\zeta}}+\gamma_{\zeta})|h_{\zeta}|^2+\gamma_{\varrho}|h_{\varrho}|^2+1},
\end{equation}
where $\zeta$, $\varrho\in \{A, B\}$ and $\zeta \neq \varrho$. We further show the expression of the transmission success probability, which will be used to optimize the average AoI. The transmission success probability is defined as the probability that the received SNR is higher than an acceptable SNR threshold $\gamma_{th}$. The exact transmission success probability expression for the two slot PNC scheme is complex for the AoI optimization. We thus adopt the asymptotic transmission success probability expression at high SNR. Based on \cite[Eq. (33)]{b19}, the asymptotic transmission success probability through $S_{\varrho}-R-S_{\zeta}$ link for the two time slot PNC scheme is given by
\begin{equation}
F_{\zeta,2TS}=1-\frac{\gamma_{th}(\gamma_{r,\zeta}+ \gamma_{\zeta}+ \gamma_{\varrho})}{\gamma_{r,\zeta}\cdot \gamma_{\varrho}}+o(\gamma_{th}^2).
\end{equation}

In the next section, we will use the above expressions to analyze and optimize the average AoI of the two time slot PNC scheme for the considered system. 

\begin{figure}
	\centering
	\includegraphics[width=0.9\linewidth]{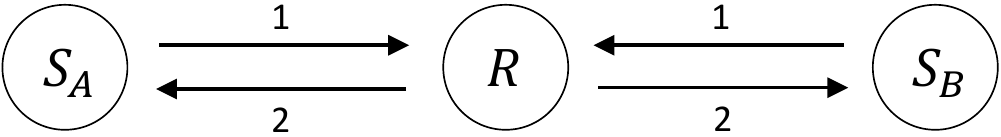}
	\caption{System model.}
	\label{fig:systemmodel}
\end{figure}

\section{Analysis and Optimization of Average AoI}
In this section, we first derive the closed-form expression of the average AoI of the considered system. Furthermore, we minimize the average AoI by optimizing the transmission power at each node under the peak power constraints.

\subsection{Analysis of Average AoI}

Since we adopt \textit{generate-at-will} model in the considered system, we assume that $S_A$ and $S_B$ generate new status updates every two time slots to keep the information at the corresponding destinations as fresh as possible. Let $U_{\zeta}(t)$ denote the generation time of the most recently received status update at $S_{\zeta}$ until time slot $t$. The AoI at time slot $t$ at $S_{\zeta}$ can then be defined as 
\begin{equation}
\Delta_{\zeta,2TS}(t)=t-U_{\zeta}(t).
\end{equation}
Let $S_{\zeta,k}$ and $T_{\zeta,k}$ denote the service time of the $k$th successfully received update at $S_{\zeta}$, and the interdeparture time between two consecutive successfully received status updates at $S_{\zeta}$, respectively. Following similar analysis to that in \cite{b21}, the average AoI at $S_{\zeta}$ can be expressed as 
\begin{equation}
\bar{\Delta}_{\zeta,2TS}=\mathbb{E}[S_{\zeta,k-1}]+\frac{\mathbb{E}\Big[T_{\zeta,k}^2\Big]}{2\mathbb{E}[T_{\zeta,k}]}-\frac{1}{2}.
\end{equation}

It is obvious that, in the considered system, the service time of each successfully received update is always a constant, which is given by $\mathbb{E}[S_{\zeta,k-1}]=2.$ As mentioned before, $S_A$ and $S_B$ are assumed to generate new status updates every time slots. Therefore, the interdeparture time $T_{\zeta,k}$ is a geometric random variable with mean $\mathbb{E}[T_{\zeta,k}]=2/F_{\zeta,2TS}$ and second moment $\mathbb{E}[T_{\zeta,k}^2]=4\cdot(2-F_{\zeta,2TS})/F_{\zeta,2TS}^2$. By substituting $\mathbb{E}[S_{\zeta,k-1}]$, $\mathbb{E}[T_{\zeta,k}]$ and $\mathbb{E}[T_{\zeta,k}^2]$ into (4), we obtain the closed-from expression of the average AoI at $S_{\zeta}$ for the two time slot PNC scheme, given by
\begin{equation}
\bar{\Delta}_{\zeta,2TS}=\frac{1}{2}+\frac{2}{F_{\zeta,2TS}}.
\end{equation}

The expected weighted sum AoI of the considered system can thus be expressed as 
\begin{equation}
\begin{aligned}
\bar{\Delta}_{2TS}&=\omega_{A}\cdot\bar{\Delta}_{A,2TS}+\omega_{B}\cdot\bar{\Delta}_{B,2TS}\\
&=\frac{1}{2}(\omega_{A}+\omega_{B})+2\left(\frac{\omega_{A}}{F_{A,2TS}}+\frac{\omega_{B}}{F_{B,2TS}}\right),
\end{aligned}
\end{equation}
where $\omega_{A}$ and $\omega_{B}$ are weighting coefficients at $S_{A}$ and $S_B$, respectively, such that $\omega_{A}+\omega_{B}=1$. 

%The detailed analysis of (3) and (4) will be shown in Section IV.

%Since we adopt generate-at-will model in the considered system, we assume that $S_A$ and $S_B$ generate new status updates every two time slots to keep the information at the corresponding destinations as fresh as possible. Under this assumption, the closed-from expression of the average AoI at $S_{\zeta}$ for the two time slot PNC scheme are given by
%\begin{equation}
%\bar{\Delta}_{\zeta,2TS}=\frac{1}{2}+\frac{2}{F_{\zeta,2TS}}.
%\end{equation}
%The expected weighted sum AoI of the considered system can thus be expressed as 
%\begin{equation}
%\begin{aligned}
%\bar{\Delta}_{2TS}&=\omega_{A}\cdot\bar{\Delta}_{A,2TS}+\omega_{B}\cdot\bar{\Delta}_{B,2TS}\\
%&=\frac{1}{2}(\omega_{A}+\omega_{B})+2\left(\frac{\omega_{A}}{F_{A,2TS}}+\frac{\omega_{B}}{F_{B,2TS}}\right),
%\end{aligned}
%\end{equation}
%where $\omega_{A}$ and $\omega_{B}$ are weighting coefficients at $S_{A}$ and $S_B$, respectively, such that $\omega_{A}+\omega_{B}=1$. The detailed analysis of (3) and (4) will be shown in Section IV.

In the next subsection, we will manage to minimize the expected weighted sum AoI by optimizing the transmission power at each node in the considered system.

\subsection{Optimization of Expected Weighted Sum AoI}
In this subsection, we optimize the expected weighted sum AoI of the considered system under the peak power constraints. By substituting (2) into (6), we can rewrite the expected weighted sum AoI of the two time slot PNC scheme as
\begin{equation}
\begin{aligned}
\bar{\Delta}_{2TS}\!=&\frac{1}{2}(\omega_{A}\!+\!\omega_{B})+2\cdot\!\left[\frac{\omega_{A}}{1\!-\!\frac{\gamma_{th}\left(\gamma_{r,A}+\gamma_{A}+\gamma_{B}\right)}{\gamma_{r,A}\cdot\gamma_{B}}}\right.\\
&\quad\qquad\qquad\ \  \qquad\qquad\ \, \left.+\frac{\omega_{B}}{1\!-\!\frac{\gamma_{th}\left(\gamma_{r,B}+\gamma_{B}+\gamma_{A}\right)}{\gamma_{r,B}\cdot\gamma_{A}}}\right]\\
=&\frac{1}{2}(\omega_{A}\!+\!\omega_{B})+2\cdot\!\left[\frac{\omega_{A}}{1\!-\!\gamma_{th}\!\left[\frac{\sigma_r^2}{P_B}\!+\!\frac{\sigma_A^2}{P_r}\!\left(\!1\!+\!\frac{P_A}{P_B}\!\right)\right]}\right.\\
&\quad\qquad\qquad\qquad\ \ \left.+\frac{\omega_{B}}{1\!-\!\gamma_{th}\!\left[\frac{\sigma_r^2}{P_A}\!+\!\frac{\sigma_B^2}{P_r}\!\left(\!1\!+\!\frac{P_B}{P_A}\!\right)\right]}\right].
\end{aligned}
\end{equation}

It is obvious that the first term on the RHS of (7) is a constant with a value of 0.5. Therefore, in order to minimize the expected weighted sum AoI of the two time slot PNC scheme, we only need to optimize the second term on the RHS of (7). We are at a point to formally formulate the following optimization problem:
%\begin{equation}
%	\bar{\Delta}_{2TS}=1+2\cdot\left(\frac{1}{F_{A,2TS}}+\frac{1}{F_{B,2TS}}\right),
%\end{equation}
%where $F_{A,2TS}=1-\gamma_{th}\Big[\frac{\sigma_r^2}{P_B}+\frac{\sigma_A^2}{P_r}\big(1+\frac{P_A}{P_B}\big)\Big]$ and $F_{B,2TS}=1-\gamma_{th}\Big[\frac{\sigma_r^2}{P_A}+\frac{\sigma_B^2}{P_r}\big(1+\frac{P_B}{P_A}\big)\Big]$.
%\begin{equation}
%\begin{aligned}
%\bar{\Delta}_{2TS}=1+&2\left[\frac{1}{1-\gamma_{th}\left[\frac{\sigma_r^2}{P_B}+\frac{\sigma_A^2}{P_r}\left(1+\frac{P_A}{P_B}\right)\right]}\right.\\
%&\quad  \left.+\frac{1}{1-\gamma_{th}\left[\frac{\sigma_r^2}{P_A}+\frac{\sigma_B^2}{P_r}\left(1+\frac{P_B}{P_A}\right)\right]}\right].
%\end{aligned}
%\end{equation}
\begin{equation}
\begin{aligned}
\min_{P_{A}, P_{B}, P_{r}} & \left\{\frac{\omega_{A}}{F_{A,2TS}}+\frac{\omega_{B}}{F_{B,2TS}}\right\}\\
\!\mbox{s.t.}\quad
&\ \, 0<P_{A}\leq P_{A}^{pk}, 0<P_{B}\leq P_{B}^{pk}, 0<P_{r}\leq P_{r}^{pk}, \\
&\ \, 0<F_{A,2TS}<1, 0<F_{B,2TS}<1,
%	&0<\gamma_{th}\left[\frac{\sigma_r^2}{P_B}+\frac{\sigma_A^2}{P_r}\left(1+\frac{P_A}{P_B}\right)\right]<1 \ {\rm \color{red}High\  SNR?}\\
%	&0<\gamma_{th}\left[\frac{\sigma_r^2}{P_A}+\frac{\sigma_B^2}{P_r}\left(1+\frac{P_B}{P_A}\right)\right]<1.
\end{aligned}
\end{equation}
%\begin{equation}
%	\begin{aligned}
%	\min_{P_{A}, P_{B}, P_{r}} \quad & \left\{\frac{1}{1-\gamma_{th}\left[\frac{\sigma_r^2}{P_B}+\frac{\sigma_A^2}{P_r}\left(1+\frac{P_A}{P_B}\right)\right]}\right.\\
%	&\quad  \left.+\frac{1}{1-\gamma_{th}\left[\frac{\sigma_r^2}{P_A}+\frac{\sigma_B^2}{P_r}\left(1+\frac{P_B}{P_A}\right)\right]}\right\}\\
%	\mbox{s.t.}\quad
%	&0<P_{A}\leq P_{A}^{pk} \\
%	&0<P_{B}\leq P_{B}^{pk} \\
%	&0<P_{r}\leq P_{r}^{pk} \\
%	&0<\gamma_{th}\left[\frac{\sigma_r^2}{P_B}+\frac{\sigma_A^2}{P_r}\left(1+\frac{P_A}{P_B}\right)\right]<1 \ {\rm \color{red}High\  SNR?}\\
%	&0<\gamma_{th}\left[\frac{\sigma_r^2}{P_A}+\frac{\sigma_B^2}{P_r}\left(1+\frac{P_B}{P_A}\right)\right]<1.
%	\end{aligned}
%\end{equation}
where $F_{A,2TS}=1-\gamma_{th}\Big[\frac{\sigma_r^2}{P_B}+\frac{\sigma_A^2}{P_r}\big(1+\frac{P_A}{P_B}\big)\Big]$ and $F_{B,2TS}=1-\gamma_{th}\Big[\frac{\sigma_r^2}{P_A}+\frac{\sigma_B^2}{P_r}\big(1+\frac{P_B}{P_A}\big)\Big]$. It is worth mentioning that the asymptotic transmission success probabilities are obtained at high SNR, which makes their values inaccurate or even less than 0 in the case of low SNR. Therefore, in order to guarantee the physical meaning of probability, we limit $0<F_{A,2TS}<1$ and $ 0<F_{B,2TS}<1$ in problem (8). Note that although $F_{A,2TS}$ and $F_{B,2TS}$ are supposed to be high probabilities, we can follow the same procedure as below to solve the optimization problem under this constraint.

%because the constraint in (6) is the most general case
%
%Without loss of generality, the asymptotic transmission success probabilities are constrained in the range of $(0,1)$ in the above optimization problem. Note that although $F_{A,2TS}$ and $F_{B,2TS}$ are obtained at high SNR, i.e., $F_{A,2TS}$ and $F_{B,2TS}$ are high probabilities, we can follow the same procedure as below to solve the optimization problem in the high SNR case.

As the objective function is a monotonously decreasing function of $P_r$, the optimal transmission power at $R$ in problem (8) is given by
\begin{equation}
P_r^*=P_r^{pk}.
\end{equation}
Thus, problem (8) is equivalent to the following problem:
\begin{equation}
\begin{aligned}
\qquad\min_{P_{A}, P_{B}} & \left\{\frac{\omega_{A}}{F_{A,2TS}'}+\frac{\omega_{B}}{F_{B,2TS}'}\right\}\\
\mbox{s.t.}\ \,
&\ \, 0<P_{A}\leq P_{A}^{pk}, 0<P_{B}\leq P_{B}^{pk} \\
&\ \, 0<F_{A,2TS}'<1, 0<F_{B,2TS}'<1,
\end{aligned}
\end{equation}
where $F_{A,2TS}'=1-\gamma_{th}\Big[\frac{\sigma_r^2}{P_B}+\frac{\sigma_A^2}{P_r^{pk}}\big(1+\frac{P_A}{P_B}\big)\Big]$ and $F_{B,2TS}'=1-\gamma_{th}\Big[\frac{\sigma_r^2}{P_A}+\frac{\sigma_B^2}{P_r^{pk}}\big(1+\frac{P_B}{P_A}\big)\Big]$. 
Problem (10) is non-convex due to the non-convexity of the object function, and is thus difficult to solve via standard convex optimization techniques \cite{b22}. In order to solve the non-convex problem, we first fix $P_A$ or $P_B$, by which we have the following lemmas:

%However, if $P_A$ or $P_B$ is fixed, it is convex in the constraint region. Therefore, using the Karush-Kuhn-Tucker (KKT) conditions \cite{b8}, which are necessary conditions of the optimal solution of problem, we have the following lemmas:

\textit{\underline{Lemma} 1:} When $P_A$ is fixed, problem (10) is converted into a convex problem, in which the optimal $P_B$ is given by
\begin{equation}
P_{B}'={\rm min}\left(P_{B}^{pk},P_{B1}\right),
\end{equation}
where $P_{B1}=\frac{\beta_B\theta_B+\varphi_B}{\kappa_B}$,
$\beta_B=P_r^{pk}\sqrt{P_A\omega_{A}\omega_{B}(P_A\sigma_A^2+P_r^{pk}\sigma_r^2)}$, 
$\theta_B=
P_A\gamma_{th}\left(\sigma_A^2+\sigma_B^2\right)+\gamma_{th}^2\sigma_r^2\left(\sigma_B^2-\sigma_A^2\right)+P_r^{pk}\gamma_{th}\sigma_r^2-P_AP_r^{pk}$,
$\varphi_B=P_AP_r^{pk}\gamma_{th}\sigma_B\left(\omega_{A}-\omega_{B}\right)\left(P_A\sigma_A^2+P_r^{pk}\sigma_r^2-\gamma_{th}\sigma_A^2\sigma_r^2\right)+\gamma_{th}^2\sigma_B\!\!\left[\omega_{B}P_A^2\sigma_A^4\!-\omega_{A}\sigma_r^4\!\left(P_r^{pk}\right)^2\!\!\!-\omega_{A}P_A\sigma_B^2\!\left(P_A\sigma_A^2\!+\!P_r^{pk}\sigma_r^2\right)\!\right]$,
$\kappa_B=\omega_{B}P_A\sigma_B\left[2P_r^{pk}\gamma_{th}\sigma_A^2-\left(P_r^{pk}\right)^2-\gamma_{th}^2\sigma_A^4\right]+\omega_A\gamma_{th}^2\sigma_B^3\left(P_A\sigma_A^2+P_r^{pk}\sigma_r^2\right)$.

\textit{Proof:} See Appendix A.

\textit{\underline{Lemma} 2:} When $P_B$ is fixed, problem (10) is converted into a convex problem, in which the optimal $P_A$ is given by
\begin{equation}
P_{A}'={\rm min}\left(P_{A}^{pk},P_{A1}\right),
\end{equation}
where $P_{A1}=\frac{\beta_A\theta_A+\varphi_A}{\kappa_A}$,
$\beta_A=P_r^{pk}\sqrt{P_B\omega_{B}\omega_{A}(P_B\sigma_B^2+P_r^{pk}\sigma_r^2)}$, 
$\theta_A=
P_B\gamma_{th}\left(\sigma_B^2+\sigma_A^2\right)+\gamma_{th}^2\sigma_r^2\left(\sigma_A^2-\sigma_B^2\right)+P_r^{pk}\gamma_{th}\sigma_r^2-P_BP_r^{pk}$,
$\varphi_A=P_BP_r^{pk}\gamma_{th}\sigma_A\left(\omega_{B}-\omega_{A}\right)\left(P_B\sigma_B^2+P_r^{pk}\sigma_r^2-\gamma_{th}\sigma_B^2\sigma_r^2\right)+\gamma_{th}^2\sigma_A\!\!\left[\omega_{A}P_B^2\sigma_B^4\!-\omega_{B}\sigma_r^4\!\left(P_r^{pk}\right)^2\!\!\!-\omega_{B}P_B\sigma_A^2\!\left(P_B\sigma_B^2\!+\!P_r^{pk}\!\sigma_r^2\right)\!\right]$,
$\kappa_A=\omega_{A}P_B\sigma_A\left[2P_r^{pk}\gamma_{th}\sigma_A^2-\left(P_r^{pk}\right)^2-\gamma_{th}^2\sigma_B^4\right]+\omega_B\gamma_{th}^2\sigma_A^3\left(P_B\sigma_B^2+P_r^{pk}\sigma_r^2\right)$.

%where $P_{A1}\!=\!\frac{\beta_A\theta_A+\varphi_A}{\kappa_A}$, $\beta_A\!=\!P_r\!\sqrt{\!P_B(P_B\sigma_B^2\!+\!P_r\sigma_r^2)}$, $\theta_A\!=$
%$\left[P_B\gamma_{th}\left(\sigma_B^2+\sigma_A^2\right)+\sigma_r^2\gamma_{th}^2\left(\sigma_A^2-\sigma_B^2\right)+P_r\gamma_{th}\sigma_r^2-P_BP_r\right]$, $\varphi_A\!=\!\gamma_{th}^2\sigma_A\left(P_B^2\sigma_A^4\!-\!P_B^2\sigma_B^2\sigma_A^2\!-\!P_r^2\sigma_r^4\!-\!P_BP_r\sigma_A^2\sigma_r^2\right)$,
%$\kappa_A=$
%$\sigma_A\!\left[P_B\gamma_{th}\sigma_B^2\left(\gamma_{th}\sigma_A^2\!-\!\gamma_{th}\sigma_B^2\!+\!2P_r\right)\!+\!P_r\gamma_{th}^2\sigma_A^2\sigma_r^2\!-\!P_BP_r^2\right]$.

\textit{Proof:} See Appendix B.

By applying the results obtained in Lemma 1 and Lemma 2, we can acquire the optimal solution to problem (10) given by the following theorem.

\textit{\underline{Theorem} 1:} The minimum expected weighted sum AoI is obtained when at least one source node transmits with the peak power. Mathematically, the optimal transmission power at the source nodes, i.e., $P_A$ and $P_B$, and the minimum expected weighted sum AoI of the two time slot PNC scheme are given as
\begin{equation}
\left(P_{A}^*,P_{B}^*\right)=\mathop{\arg\min}_{(P_A,P_B)\in\left\{\left(P_A^{pk},P_B'\right),\left(P_A',P_{B}^{pk}\right)\right\}}\left\{f(P_A,P_B)\right\},
\end{equation}
\begin{equation}
\bar{\Delta}_{2TS}^*=\frac{1}{2}+f\left(P_A^*,P_B^*\right),
\end{equation}
where $f(P_A,P_B)$ is the objective function in problem (10).

\textit{Proof:} See Appendix C.

\section{Simulation Results and Discussions}

In this section, we present simulation results to validate our analytical results. Without loss of generality, we consider that the two source nodes are of the same importance in the system, i.e., $\omega_{A}=\omega_{B}=0.5$. In the following simulations, it is also assumed that $P_r^*=P_r^{pk}=0.75P_A^{pk}$, $\sigma_A^2=\sigma_B^2=\sigma_r^2=0.001$ and $\gamma_{th}=20$dB. For the exact analytical results shown in Fig. 2, we plot them by substituting the exact transmission success probability expression given in \cite[Eq. (29)]{b19} into our AoI analysis.
Each simulation curve presented in the section is averaged over $10^7$ time slots.

We first plot the expected weighted sum AoI of the two time slot PNC scheme against the transmission power at $S_B$ (i.e., $P_B$) by fixing the transmission power at $S_A$ (i.e., $P_A$). As shown in Fig. 2, we consider two cases with $P_A$ being 1 and 1.5, respectively. In both cases, it is apparent that the expected weighted sum AoI first decreases and then increases as $P_B$ increases. This is because when $S_B$ transmits with higher power, although $S_A$ will receive better signals, the received SNR at $S_B$ will be lower. The resulting transmission success probabilities through the $S_B-R-S_A$ link and the $S_A-R-S_B$ link will thus increase and decrease, respectively. However, no matter how high $F_{A,2TS}'$ is, the minimum average AoI at $S_A$ can only be reduced to 2, while the average AoI at $S_B$ will grow rapidly when $F_{B,2TS}'$ is low. Therefore, to high transmission power at $S_B$ in turn increases the expected weighted sum AoI of the considered system. It is also worth mentioning that the optimal transmission power at $S_B$ to minimize the expected weighted sum AoI in the case of $P_A=1$ and $P_A=1.5$ are $P_B=1.196$ and $P_B=1.718$, respectively, which are consistent with the results calculated by using the formula in Lemma 2. We can also find that, in both cases, the asymptotic analytical result is close to the simulation result and the exact analytical result when $P_B \in (0.75,2)$. Specifically, in this region, when $P_A=1$ and $P_A=1.5$, we can have $F_{A,2TS}'>0.56$, $F_{B,2TS}'>0.50$ and $F_{A,2TS}'>0.60$, $F_{B,2TS}'>0.72$, respectively. This observation validates the rationality of using the asymptotic result at high SNR, or high transmission success probability in our analysis.

After verifying the tightness between the asymptotic expected weighted sum AoI and the exact values at high transmission success probabilities, we now investigate the optimal operating point of the considered system with the help of the asymptotic expression. In Fig. 3, we report the expected weighted sum AoI of the system in function of $P_A$ and $P_B$, whose peak values are assumed to be $P_A^{pk}=1$ and $P_B^{pk}=2$, respectively. In order to ensure the tightness, we also assume that $0.5<F_{A,2TS}'<1$ and $0.5<F_{B,2TS}'<1$.
The first thing we can see is that if both $P_A$ and $P_B$ approach 0.4, the expected weighted sum AoI is high as the received SNR of both links has a high probability of being below the threshold. We can also observe that if both $S_A$ and $S_B$ transmits with the peak power, the resulting expected weighted sum AoI is not minimum in the considered system, which is consistent with what we observed when $P_A=1$ in Fig. 2. It is also worth mentioning that the optimal expected weighted sum age of the considered system (marked in red in Fig. 3) is achieved for $P_A^*=1$, $P_B^*=1.196$ and $\bar{\Delta}_{2TS}^*=3.636$, which validates our results in theorem 1.

\begin{figure}
	\centering
	\includegraphics[width=0.95\linewidth]{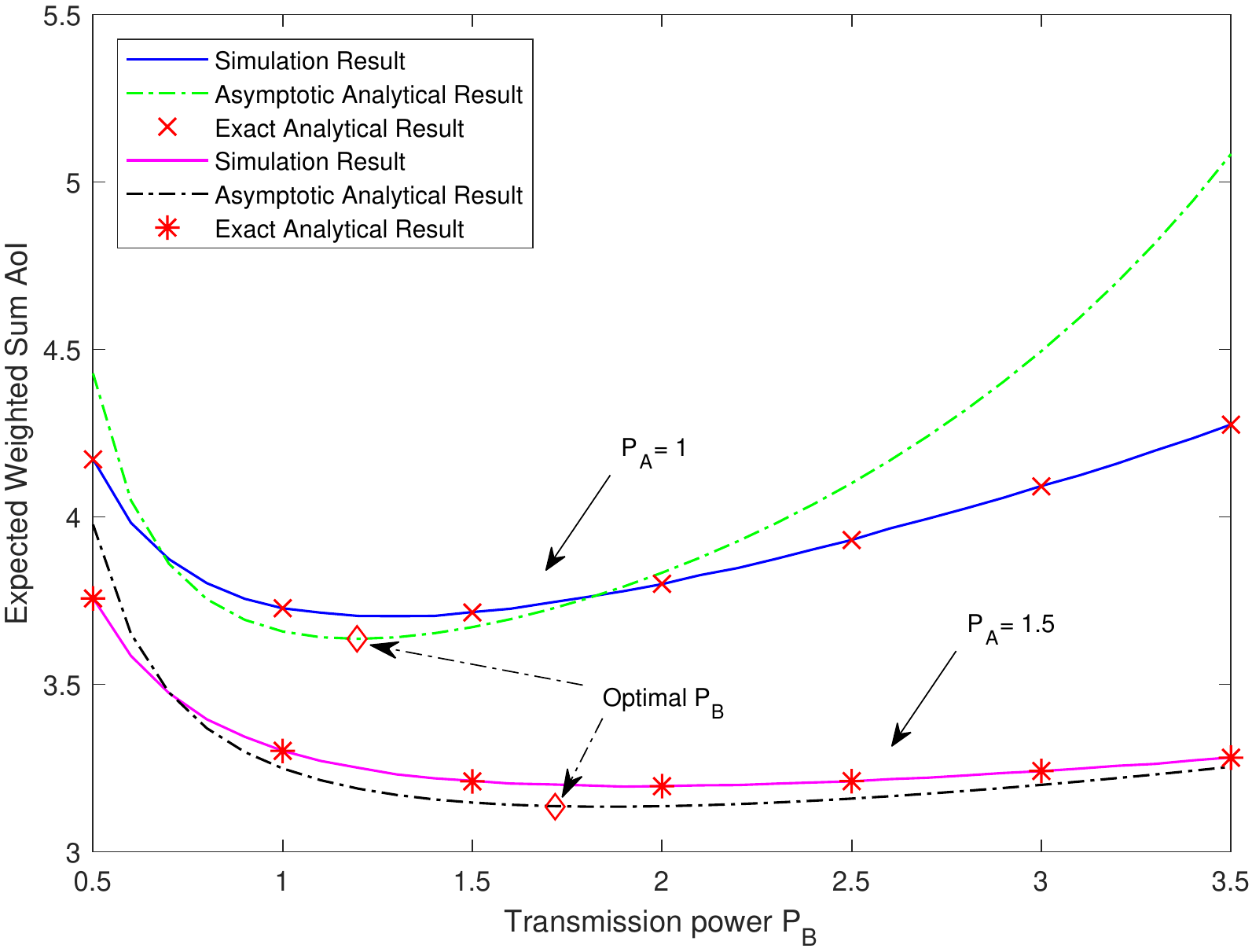}
	\caption{{Expected weighted sum AoI of the considered system versus the transmission power $P_B$ for different transmission power $P_A$.}}
	\label{}
\end{figure}

\begin{figure}
	\centering
	\includegraphics[width=0.97\linewidth]{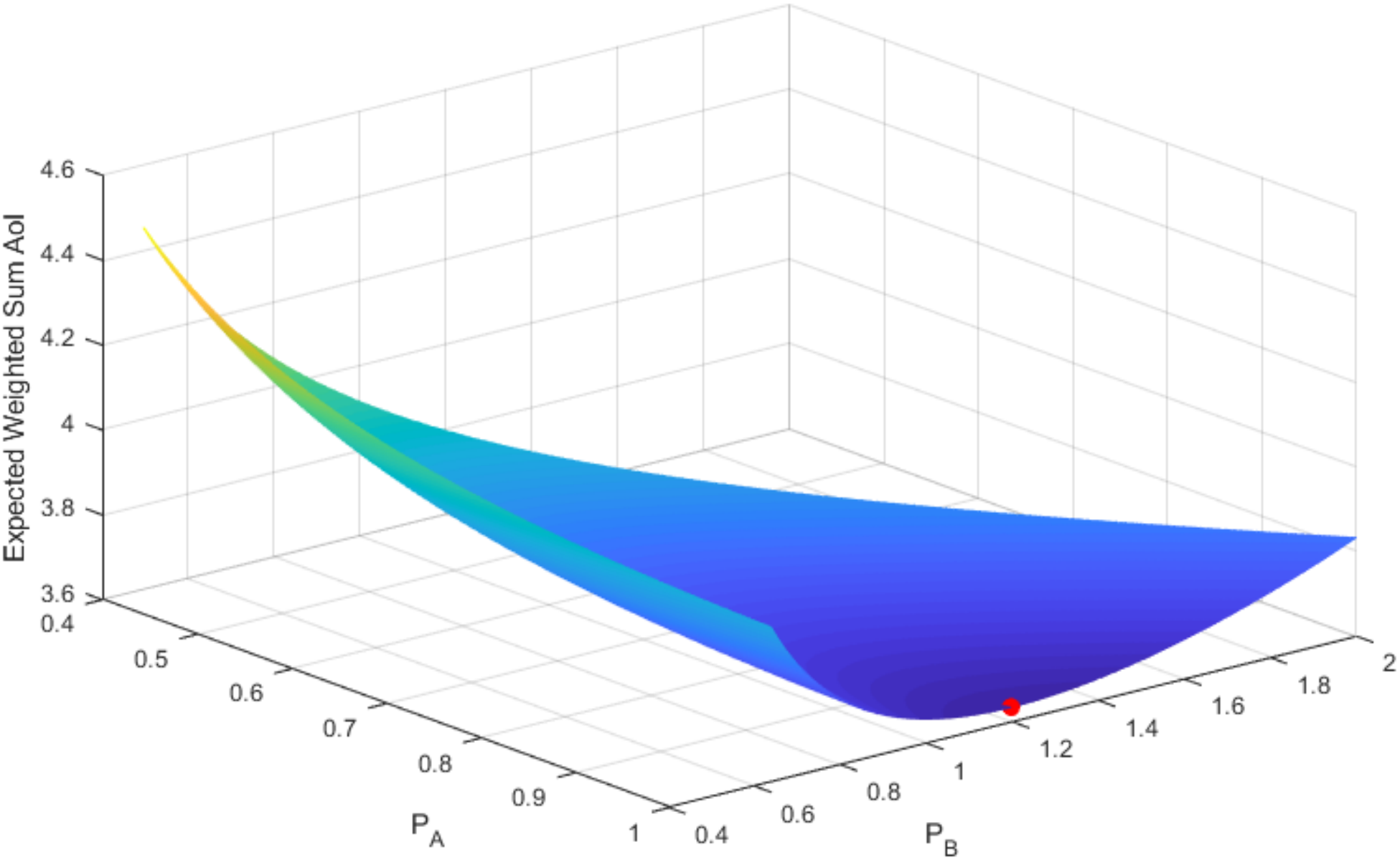}
	\caption{Expected weighted sum AoI of the considered system in function of $P_A$ and $P_B$ at high transmission success probabilities.}
	\label{}
\end{figure}

\section{Conclusions}
In this paper, we analyzed and optimized the average age of information (AoI) of a two-way cooperative status update system, where two source nodes timely exchange status updates with each other with the help of a relay. By analyzing the evolution of the AoI, we derived a closed-form expression of the expected weighted sum AoI for the considered system as a function of the signal-to-noise ratio (SNR) at each node. Under the peak power constraints, we further figured out the optimal transmission power at each node that minimizes the expected weighted sum AoI. We found that when the AoI of the considered system is minimum, at least one source node transmits with its peak power, while the optimal transmission power of the other source node may be lower than than the peak value. Simulation results validated our theoretical analysis.

\section*{Appendix}
\textit{A. Proof of Lemma 1}

When $P_A$ is fixed, the objective function in problem (10) is a function only of $P_B$, i.e. $f(P_B)$. To verify its convexity, we derive the second-order derivative of $f(P_B)$ with respect to (w.r.t.) $P_B$. After some algebra manipulations, we have
\begin{equation}
\begin{aligned}
	\frac{\partial^2 f(P_B)}{\partial P_B^2}=
	&\frac{2\omega_{A}\gamma_{th}\left(P_A\sigma_A^2+P_r^{pk}\sigma_r^2\right)\left(P_r^{pk}-\gamma_{th}\sigma_A^2\right)}{P_B^3\left(P_r^{pk}\right)^2\left(F_{A,2TS}'\right)^3}\\
	&+\frac{2\omega_{B}\gamma_{th}^2\sigma_B^4}{P_A^2\left(P_r^{pk}\right)^2\left(F_{B,2TS}'\right)^3}.
\end{aligned}
\end{equation}

As $0<F_{A,2TS}'<1$ and $0<F_{B,2TS}'<1$, the denominators of the first and second terms on the RHS of (15) are always larger than zero. Based on $F_{A,2TS}'<1$, after some algebra manipulations, we have that $P_r^{pk}-\gamma_{th}\sigma_A^2>0$. Therefore, we always have $\frac{\partial^2 f(P_B)}{\partial P_B^2}>0$, which indicates that when $P_A$ is fixed, the objective function in problem (10) is convex in the constraint region.

In order to find the optimal $P_B$, which minimizes the convex function $f(P_B)$, we need to calculate the extreme point of the function. Specifically, we need to solve $\frac{\partial f(P_B)}{\partial P_B}=0$, whose solutions are given by 
\begin{equation}
P_{B1}=\frac{\beta_B\theta_B+\varphi_B}{\kappa_B} \quad  {\rm and} \quad  P_{B2}=\frac{\beta_B'\theta_B'+\varphi_B'}{\kappa_B'}, 
\end{equation}
where  $\beta_B=\beta_B'=P_r^{pk}\sqrt{P_A\omega_{A}\omega_{B}(P_A\sigma_A^2+P_r^{pk}\sigma_r^2)}$,
$\theta_B=-\theta_B'=
P_A\gamma_{th}\left(\sigma_A^2+\sigma_B^2\right)\!+\gamma_{th}^2\sigma_r^2\left(\sigma_B^2-\sigma_A^2\right)+$
$P_r^{pk}\gamma_{th}\sigma_r^2-P_AP_r^{pk}$,
$\varphi_B=\varphi_B'=P_AP_r^{pk}\gamma_{th}\sigma_B\left(\omega_{A}-\omega_{B}\right)\left(P_A\sigma_A^2+P_r^{pk}\sigma_r^2-\gamma_{th}\sigma_A^2\sigma_r^2\right)+\gamma_{th}^2\sigma_B\!\!\left[\omega_{B}P_A^2\sigma_A^4\!-\omega_{A}\sigma_r^4\!\left(P_r^{pk}\right)^2\!\!\!-\omega_{A}P_A\sigma_B^2\!\left(P_A\sigma_A^2\!+\!P_r^{pk}\sigma_r^2\right)\!\right]$,
$\kappa_B=\kappa_B'=\omega_{B}P_A\sigma_B\left[2P_r^{pk}\gamma_{th}\sigma_A^2-\left(P_r^{pk}\right)^2-\gamma_{th}^2\sigma_A^4\right]+\omega_A\gamma_{th}^2\sigma_B^3\left(P_A\sigma_A^2+P_r^{pk}\sigma_r^2\right)$.

%$\theta_B\!=\!-\theta_B'\!=\!
%P_A\gamma_{th}\!\left(\sigma_A^2\!+\!\sigma_B^2\right)\!+\!\gamma_{th}^2\sigma_r^2\!\left(\sigma_B^2\!-\!\sigma_A^2\right)\!+\!P_r\gamma_{th}\sigma_r^2\!-\!P_AP_r$.

%$\beta_B\!=\!\beta_B'\!=\!P_r^{pk}\sqrt{\!P_A\omega_{A}\omega_{B}(P_A\sigma_A^2\!+\!P_r^{pk}\sigma_r^2)}$, 

%$\theta_B\!=\!-\theta_B'\!=\!
%P_A\gamma_{th}\!\left(\sigma_A^2\!+\!\sigma_B^2\right)\!+\!\gamma_{th}^2\sigma_r^2\!\left(\sigma_B^2\!-\!\sigma_A^2\right)\!+\!P_r\gamma_{th}\sigma_r^2\!-\!P_AP_r$,

%$\varphi_B=\!\varphi_B'\!=\!\gamma_{th}^2\sigma_B\left[\omega_{B}P_A^2\sigma_A^4-\omega_{A}P_r^2\sigma_r^4-\omega_{A}P_A\sigma_B^2\left(P_A\sigma_A^2+P_r\sigma_r^2\right)\right]+P_AP_r\gamma_{th}\sigma_B\left(\omega_{A}-\omega_{B}\right)\left(P_A\sigma_A^2+P_r\sigma_r^2-\gamma_{th}\sigma_A^2\sigma_r^2\right)$,

%$\kappa_B\!=\!\kappa_B'\!=\omega_{B}P_A\sigma_B\!\left(2P_r\gamma_{th}\sigma_A^2\!\!-\!P_r^2\!-\!\gamma_{th}^2\sigma_A^4\right)+\omega_A\gamma_{th}^2\sigma_B^3\!\left(P_A\sigma_A^2\!\!+\!\!P_r\sigma_r^2\right)$

However, due to the convexity of $f(P_B)$, only one of the above solutions is in the feasible region. Therefore, in the following, we further verify the correctness of the above two solutions. With the help of $F_{A,2TS}'<1$ and $F_{B,2TS}'<1$, we have
\begin{equation}
	g_B<P_B<h_B,
\end{equation}
where $g_B\!=\!\frac{\gamma_{th}\left(\!P_r^{pk}\sigma_r^2+P_A\sigma_A^2\!\right)}{P_r^{pk}-\gamma_{th}\sigma_A^2}$ and $h_B\!=\!\frac{P_AP_r-\gamma_{th}\left(\!P_A\sigma_B^2+P_r\sigma_r^2\!\right)}{\gamma_{th}\sigma_B^2}$. Substituting $P_{B2}$ into (17), after some algebra manipulations, the differences between $P_{B2}$ and the boundary points are given by 
\begin{equation}
	P_{B2}-g_B=\frac{-P_r^{pk}X_gY_g}{W_gZ_g} \quad {\rm and} \quad P_{B2}-h_B=\frac{-P_r^{pk}X_hY_h}{W_hZ_h},
\end{equation}
where 
$W_g=\sigma_B\left(P_r^{pk}-\gamma_{th}\sigma_A^2\right)$, $W_h=\gamma_{th}\sigma_B^2$, $X_g=\left(P_r^{pk}-\gamma_{th}\sigma_A^2\right)\beta_B+\omega_{A}\left(P_A\gamma_{th}\sigma_A^2\sigma_B+P_r^{pk}\gamma_{th}\sigma_B^2\sigma_r\right)$,
$X_h=\gamma_{th}\sigma_B\beta_B+\omega_{B}P_A\left(P_r^{pk}-\gamma_{th}\sigma_A^2\right)$,
$Y_g=Y_h=\theta_B$ and $Z_g=Z_h=\kappa_B$. 
Since we have $P_r^{pk}-\gamma_{th}\sigma_A^2>0$, it is obvious that $X_g>0$, $X_h>0$, $W_g>0$ and $W_h>0$. In order to investigate whether $P_{B2}$ is in the constraint region, we need to consider the signs of $Y_g=Y_h$ and $Z_g=Z_h$ in the following.

Based on (17), we should always have
\begin{small}
	\begin{equation}
	g_B\!=\!\frac{\gamma_{th}\!\left(P_r^{pk}\sigma_r^2+P_A\sigma_A^2\right)}{P_r^{pk}-\gamma_{th}\sigma_A^2}\!<\!h_B\!=\!\frac{P_AP_r-\gamma_{th}\!\left(P_A\sigma_B^2+P_r\sigma_r^2\right)}{\gamma_{th}\sigma_B^2}.
	\end{equation}
\end{small} 
$\!\!$After some algebraic manipulation on (19), we can readily find that $Y_g=Y_h<0$. We now investigate $P_{B2}$ for the following two possible cases:

$1)$ If $Z_g=Z_h<0$, we have $P_{B2}-g_B>0$ and $P_{B2}-h_B>0$, i.e., $g_B<h_B<P_{B2}$, which contradicts the feasible region. 

$2)$ If $Z_g=Z_h>0$, we have $P_{B2}-g_B<0$ and $P_{B2}-h_B<0$, i.e., $P_{B2}<g_B<h_B$, which contradicts the feasible region. 

\noindent From the above two cases, it is obvious that $P_{B2}$ is not valid. 

In a similar manner, we can also prove that $P_{B1}$ is within the feasible region, which indicates that $P_{B1}$ is the only valid solution to minimize $f(P_B)$. Since it has been shown that $f(P_B)$ is a convex function when $P_A$ is fixed, on the left hand side (LHS) of $P_{B1}$, $f(P_B)$ decreases monotonically as $P_B$ increases. Therefore, if $P_{B1}>P_{B}^{pk}$, the minimum value of $f(P_B)$ is obtained at $P_{B}^{pk}$. This completes the proof.

\textit{B. Proof of Lemma 2}

%{\color{red}Lemma 2 can be proved in a similar manner to Appendix A and is thus omitted for brevity.} 

When $P_B$ is fixed, the objective function in problem (10) is a function only of $P_A$, i.e., $f(P_A)$. To verify its convexity, we derive the second-order derivative of $f(P_A)$ w.r.t. $P_A$, given by
\begin{equation}
\begin{aligned}
\frac{\partial^2 f(P_A)}{\partial P_A^2}=
&\frac{2\omega_{B}\gamma_{th}\left(P_B\sigma_B^2+P_r^{pk}\sigma_r^2\right)\left(P_r^{pk}-\gamma_{th}\sigma_B^2\right)}{P_A^3\left(P_r^{pk}\right)^2\left(F_{B,2TS}'\right)^3}\\
&+\frac{2\omega_{A}\gamma_{th}^2\sigma_A^4}{P_B^2\left(P_r^{pk}\right)^2\left(F_{A,2TS}'\right)^3}.
\end{aligned}
\end{equation}

Based on $F_{B,2TS}'<1$, after some algebra manipulations, we have that $P_r^{pk}-\gamma_{th}\sigma_B^2>0$. According to the analysis of (15) in Appendix A, we can have $\frac{\partial^2 f(P_A)}{\partial P_A^2}>0$, which indicates that when $P_B$ is fixed, the objective function in problem (10) is convex in the constraint region. In addition, the optimal $P_A$ to minimize $f(P_A)$ are given by
\begin{equation}
P_{A1}=\frac{\beta_A\theta_A+\varphi_A}{\kappa_A} \quad  {\rm and} \quad  P_{A2}=\frac{\beta_A'\theta_A'+\varphi_A'}{\kappa_A'}, 
\end{equation}
where  $\beta_A=\beta_A'=P_r^{pk}\sqrt{P_B\omega_{B}\omega_{A}(P_B\sigma_B^2+P_r^{pk}\sigma_r^2)}$,
$\theta_A=-\theta_A'=
P_B\gamma_{th}\left(\sigma_B^2+\sigma_A^2\right)\!+\gamma_{th}^2\sigma_r^2\left(\sigma_A^2-\sigma_B^2\right)+$
$P_r^{pk}\gamma_{th}\sigma_r^2-P_BP_r^{pk}$,
$\varphi_A=\varphi_A'=P_BP_r^{pk}\gamma_{th}\sigma_A\left(\omega_{B}-\omega_{A}\right)\left(P_B\sigma_B^2+P_r^{pk}\sigma_r^2-\gamma_{th}\sigma_B^2\sigma_r^2\right)+\gamma_{th}^2\sigma_A\!\!\left[\omega_{A}P_B^2\sigma_B^4\!-\omega_{B}\sigma_r^4\!\left(P_r^{pk}\right)^2\!\!\!-\omega_{B}P_B\sigma_A^2\!\left(P_B\sigma_B^2\!+\!P_r^{pk}\!\sigma_r^2\right)\!\right]$,
$\kappa_A=\kappa_A'=\omega_{A}P_B\sigma_A\left[2P_r^{pk}\gamma_{th}\sigma_B^2-\left(P_r^{pk}\right)^2-\gamma_{th}^2\sigma_B^4\right]+\omega_B\gamma_{th}^2\sigma_A^3\left(P_B\sigma_B^2+P_r^{pk}\sigma_r^2\right)$.

Similar to Appendix A, we can prove that $P_{A1}$ is the only valid solution to minimize $f(P_A)$. Recall that $f(P_A)$ is a convex function when $P_B$ is fixed. This means that, on the LHS of $P_{A1}$, $f(P_A)$ is a monotonically decreasing function. Therefore, if $P_{A1}>P_{A}^{pk}$, the minimum value of $f(P_A)$ is obtained at $P_{A}^{pk}$. This completes the proof.

\textit{C. Proof of Theorem 1}

Recall that the objective function in problem (10) is given by
\begin{equation}
\begin{aligned}
f(P_A,P_B)&=\frac{\omega_{A}}{1-\gamma_{th}\Big[\frac{\sigma_r^2}{P_B}+\frac{\sigma_A^2}{P_r^{pk}}\big(1+\frac{P_A}{P_B}\big)\Big]}\\
&\qquad\quad +\frac{\omega_{B}}{1-\gamma_{th}\Big[\frac{\sigma_r^2}{P_A}+\frac{\sigma_B^2}{P_r^{pk}}\big(1+\frac{P_B}{P_A}\big)\Big]}.
\end{aligned}	
\end{equation}
Suppose when $\frac{\widetilde{P}_{A}}{\widetilde{P}_{B}}=\rho$, $\widetilde{P}_{A}<P_{A}^{pk}$ and $\widetilde{P}_{B}<P_{B}^{pk}$, the corresponding $f(\widetilde{P}_{A},\widetilde{P}_{B})$ is minimized. We can always have (a) $\frac{P_{A}^{pk}}{\widetilde{P}_{B}'}=\rho$, where $P_{A}^{pk}>P_A$ and $\widetilde{P}_{B}'>\widetilde{P}_{B}$; (b) $\frac{\widetilde{P}_{A}'}{P_{B}^{pk}}=\rho$, where $\widetilde{P}_{A}'>\widetilde{P}_{A}$ and $P_{B}^{pk}>P_B$. In the following, we further compare $f(P_{A}^{pk}, \widetilde{P}_{B}')$ and $f(\widetilde{P}_{A}', P_{B}^{pk})$ with $f(\widetilde{P}_{A}, \widetilde{P}_{B})$, respectively.

To investigate $f(P_{A}^{pk}, \widetilde{P}_{B}')$, we now consider $\widetilde{P}_{B}'$ for the following two cases:

1) If $\widetilde{P}_{B}'<P_{B}^{pk}$, $\widetilde{P}_{B}'$ is in the feasible region. Based on (22), when $\frac{P_{A}^{pk}}{\widetilde{P}_{B}'}=\frac{\widetilde{P}_{A}}{\widetilde{P}_{B}}=\rho$, $P_{A}^{pk}>\widetilde{P}_{A}$ and $\widetilde{P}_{B}'>\widetilde{P}_{B}$, it is obvious that $f(P_{A}^{pk}, \widetilde{P}_{B}')<f(\widetilde{P}_{A},\widetilde{P}_{B})$.

2) If $\widetilde{P}_{B}'>P_{B}^{pk}$, $\widetilde{P}_{B}'$ is not in the feasible region. However, we now have $\frac{P_{A}^{pk}}{P_{B}^{pk}}>\frac{P_{A}^{pk}}{\widetilde{P}_{B}'}=\rho$. Therefore, we can always find $\widetilde{P}_{A}'$, which follows $\widetilde{P}_{A}'<P_{A}^{pk}$, so that $\frac{\widetilde{P}_{A}'}{P_{B}^{pk}}=\rho$. As $\frac{\widetilde{P}_{A}'}{P_{B}^{pk}}=\frac{\widetilde{P}_{A}}{\widetilde{P}_{B}}=\rho$ and $P_{B}^{pk}>\widetilde{P}_{B}$, it is obvious that $\widetilde{P}_{A}'>\widetilde{P}_{A}$. Therefore, we can readily have that $f(\widetilde{P}_{A}', P_{B}^{pk})<f(\widetilde{P}_{A},\widetilde{P}_{B})$.

Similarly, when investigating $f(\widetilde{P}_{A}', P_{B}^{pk})$, we have the following two cases:

1) If $\widetilde{P}_{A}'<P_{A}^{pk}$, $f(\widetilde{P}_{A}', P_{B}^{pk})<f(\widetilde{P}_{A},\widetilde{P}_{B})$. 

2) If $\widetilde{P}_{A}'>P_{A}^{pk}$, $f(P_{A}^{pk}, \widetilde{P}_{B}')<f(\widetilde{P}_{A},\widetilde{P}_{B})$.

Based on the above cases, we can always find feasible $\widetilde{P}_{A}'$ and $\widetilde{P}_{B}'$ to ensure that $f(\widetilde{P}_{A}', P_{B}^{pk})<f(\widetilde{P}_{A},\widetilde{P}_{B})$ and $f(P_{A}^{pk}, \widetilde{P}_{B}')<f(\widetilde{P}_{A},\widetilde{P}_{B})$, which contradicts our hypothesis. It indicates that at least one of $P_A$ and $P_B$ should be chosen as the corresponding peak value to minimize the objective function $f(P_A, P_B)$. By applying the results obtained in Lemma 1 and Lemma 2, we complete the proof of the theorem.

\end{document}